\documentclass[12pt,a4paper]{article}

\pdfoutput=1

\usepackage{jheppub}

\title{ ${\cal N}=2$ gauge theories and quantum phases}

\author{Jorge G. Russo}
\affiliation{Instituci\'o Catalana de Recerca i Estudis Avan\c cats (ICREA),\\
Pg. Lluis Companys 23, 08010 Barcelona, Spain.}
\affiliation{ECM Department and Institute for Sciences of the Cosmos, Facultat de F\'\i sica, \\
Universitat de Barcelona, 
Mart\'i Franqu\`es 1, E08028 Barcelona, Spain.}
\emailAdd{jorge.russo@icrea.cat}

\abstract{
The  partition function of general  ${\cal N}=2$ supersymmetric $SU(2)$ Yang-Mills theories on a four-sphere localizes to a matrix integral. 
We show that  in the decompactification limit,  and in a certain regime,  the integral is dominated by a saddle point. When this takes effect,
the free energy is exactly given in terms of the prepotential, $F=-R^2{\rm Re} (4\pi i {\cal F}) $, evaluated at the singularity of the Seiberg-Witten curve where the dual magnetic variable
$a_D$ vanishes.
We also show that the superconformal fixed point of  massive supersymmetric QCD with gauge group $SU(2)$ is associated with the existence of a quantum phase transition.
Finally, we discuss the case of ${\cal N}=2^*$ $SU(2)$  Yang-Mills theory and show that
the theory does not exhibit phase transitions.
 }

\usepackage{color}

\begin{document}

\makebox[0pt][l]{\hspace*{120mm} \parbox{3cm}{ICCUB-14-065}}
\hspace*{-16mm}
\maketitle


\def\Xint#1{\mathchoice
   {\XXint\displaystyle\textstyle{#1}}%
   {\XXint\textstyle\scriptstyle{#1}}%
   {\XXint\scriptstyle\scriptscriptstyle{#1}}%
   {\XXint\scriptscriptstyle\scriptscriptstyle{#1}}%
   \!\int}
\def\XXint#1#2#3{{\setbox0=\hbox{$#1{#2#3}{\int}$}
     \vcenter{\hbox{$#2#3$}}\kern-.52\wd0}}
\def\ddashint{\Xint=}
\def\dashint{\Xint-}

\newcommand{\be}{\begin{equation}}\newcommand{\ee}{\end{equation}}
\newcommand{\bea}{\begin{eqnarray}} \newcommand{\eea}{\end{eqnarray}}
\def\sech{ {\rm sech}}
\def\p{\partial}
\def\pa{\partial}
\def\ov{\over }
\def\a{\alpha }
\def\g{\gamma}
\def\s{\sigma }
\def\td{\tilde }
\def\vp{\varphi}
\def\gd{\nu }
\def \ha {{1 \over 2}}

\def\KK{{\cal K}}

\newcommand\cev[1]{\overleftarrow{#1}}


\section{Introduction}

Non-abelian gauge theories exhibit a vast number of extremely interesting phenomena.
Many of these involve non-perturbative physics and are also present even in ${\cal N}=2$ supersymmetric gauge theories.
These theories can be viewed as a precise laboratory to test our intuition on the dynamics of quantum chromodynamics in terms of exact results, that include all perturbative and non-perturbative
contributions in terms of closed, analytic formulas.
Since the pioneering work of Seiberg and Witten \cite{Seiberg:1994rs,Seiberg:1994aj}, in the last two decades
there were   numerous  remarkable discoveries and  key constructions
(for reviews, see e.g. \cite{D'Hoker:1999ft,Tachikawa:2013kta}).

Using holomorphy, the approach of Seiberg and Witten \cite{Seiberg:1994rs,Seiberg:1994aj}
leads to the exact computation of the low-energy effective action for  general ${\cal N}=2$ supersymmetric gauge theories.
A systematic way to obtain these solutions by means of instanton counting was found some years later by Nekrasov \cite{Nekrasov:2002qd,Nekrasov:2003rj}.
Another approach is supersymmetric localization, which was used to determine 1/2 supersymmetric observables in  ${\cal N}=2$  theories on ${\mathbb S}^4$, such as the partition function and the circular Wilson loop,
in terms of an $r$-dimensional integral, where $r$ is the rank of the gauge group.
For   $SU(2)$ gauge groups, localization  thus reduces the computation of the exact partition function to a single integral,
which means an enormous simplification as compared with the original infinite dimensional
functional integral.
However, the partition function  is still very difficult to compute exactly
because the integrand is complicated, involving Barnes G-functions and the instanton factor.


One case that can be computed exactly is  $SU(N)$ ${\cal N}=2$ supersymmetric gauge theories in the limit of large $N$. Taking the infinite $N$ limit leads to two important simplifications. First, instanton contributions are exponentially suppressed with $N$ and
as a result the instanton factor is set to 1. Secondly, in this limit the integral is determined by
a saddle point. In turn, this permits to  calculate the partition function and Wilson loop 
exactly as a function of the coupling in terms of analytic formulas.
Along the way one obtains predictions for  theories with AdS/CFT duals.
Using these ideas, many new insights into the  physics of  large $N$ four-dimensional  gauge theories 
have recently been obtained \cite{Rey:2010ry,Passerini:2011fe,Russo:2012ay,Buchel:2013id,Russo:2013qaa,
Russo:2013kea,Bobev:2013cja,Russo:2013sba,Bigazzi:2013xia,Billo:2014bja,Chen:2014vka,Marmiroli:2014ssa,Zarembo:2014ooa}.

One of the surprising outcomes of these studies  is the proliferation  of large $N$ quantum phase transitions in the decompactification limit, which
 seem to be generic features of  ${\cal N}=2$ theories with massive matter
(exceptions include massive deformations of the ${\cal N}=2$ superconformal theory \cite{Russo:2013kea}).
The quantum critical points originate from resonances that appear whenever the coupling is such that the saddle point hits points in the Coulomb branch of the moduli space where there are massless excitations. 
In some cases, this effect leads to complicated phase structures.

One of the motivations of this paper is to elaborate on the bridge between  these recent
results from localization  and the extense literature based on the
Seiberg and Witten solution.
While the phase transitions were detected at large $N$, it is plausible that even for low-rank
gauge groups, such as $SU(2)$, there might be non-analytic features in the free energy
due to the fact that,  for certain critical couplings,
configurations crossing massless singularities may dominate the integral.

In this paper we will focus on two supersymmetric gauge theories where the Seiberg-Witten solution has been extensively studied: the $SU(2)$ super QCD  with $N_f=2$ massive multiplets, and the $SU(2)$ ${\cal N}=2^*$ theory, corresponding to a massive deformation of ${\cal N}=4$ super Yang-Mills.
In the large $N$ $SU(N)$ version, these two theories exhibit  phase transitions at certain couplings.
For the SQCD theory, there are two phases \cite{Russo:2013kea,Russo:2013sba}, the weak coupling phase with $2\Lambda <M $
and the strong coupling phase with $2\Lambda >M $.
On the other hand, ${\cal N}=2^*$ theory exhibits an infinite number of phase transitions
undergoing as the coupling $\lambda $ is increased and accumulating at $\lambda =\infty $ with the asymptotic critical coupling $\sqrt{\lambda }\sim n\pi$, where $n\gg1 $ is an integer \cite{Russo:2013qaa,Russo:2013kea,Chen:2014vka,Zarembo:2014ooa}.

The basic starting point will be the observation that, in a certain regime of the coupling, taking the decompactification limit in localization
formulas permits to write the Pestun partition function in terms of the Seiberg-Witten prepotential.
The simplification occurs provided  a saddle point exists at sufficiently large 
 value of the integration variable.

In compactifying the gauge theories on ${\mathbb S}^4$, the curvature couplings generate a scalar potential that lifts the  vacuum degeneracy.
In \cite{Russo:2013sba} it was pointed out that 
sending subsequently the radius of the sphere to infinity defines a unique vacuum in the decompactification limit, in much the same way as switching on a small external magnetic field in a Heisenberg ferromagnet selects a unique vacuum.
The present results show that this ``$S^4$ vacuum" corresponds to minimizing the Seiberg-Witten prepotential. The curvature couplings indeed do not drop out in the infinite radius limit,
but  contribute to the Seiberg-Witten prepotential in the classical or one-loop term,  as we shall explain.

We will begin with the simplest example, pure $SU(2)$ super Yang-Mills theory.


\section{Pure $SU(2)$ Super Yang-Mills theory}

\subsection{The Seiberg-Witten solution}

The supersymmetric vacuum in pure ${\cal N}=2$ $SU(N)$ gauge theories is characterized by
the expectation value of the scalar field of the vector multiplet, given by
\be
\label{vac}
\Phi ={\rm diag} (a_1,...,a_N)\ ,\qquad \sum_{i=1}^N a_i =0\ .
\ee
The low-energy effective action in ${\cal N}=2$ gauge theory is fully determined in terms of the prepotential  ${\cal F}(a_i)$. 
The magnetic dual variables  defined by
\be
a_{Di} = \frac{\partial {\cal F} }{\partial a_i}\ ,
\ee
will play an important r\^ ole in what follows.

Our  discussion will be restricted to  $SU(2)$  gauge group. In this case the coupling constant is given by
\be
\tau (a)=\frac{\partial^2 {\cal F} }{\partial a^2}\ .
\ee
It represents  the renormalized  coupling in the vacuum (\ref{vac}),
\bea
\tau(a) &=& 2\tau_{\rm UV} - \frac{8}{2\pi i}\ln \frac{a}{\Lambda_{\rm UV}}+... 
\noindent\\
&=&  - \frac{8}{2\pi i}\ln \frac{a}{\Lambda}+... 
\label{aras}
\eea
where $\Lambda $ is the dynamical scale, related to the renormalization scale $\Lambda_{\rm UV} $ by
\be
\Lambda = \Lambda_{\rm UV} \ e^{\frac{1}{2} \pi i\tau_{\rm UV}}\ .
\ee
Equation (\ref{aras}) shows the one-loop contribution to $\tau(a)$.
The exact expression for the coupling at a given vacuum parametrized by $a$ is obtained from
the Seiberg-Witten (SW) solution.
For pure $SU(2)$ SYM, the SW curve is given by \cite{Seiberg:1994rs}
\be
\label{curr}
y^2 = (x^2-\Lambda^4) (x-u)\ ,
\ee
where $u$ is the  gauge invariant parameter
\be
u=  \langle {\rm tr}\, \Phi^2 \rangle =2 a^2+...\ ,
\ee
and dots stand for quantum corrections.
The  curve  (\ref{curr}) has singularities at $u=\pm \Lambda^2 $.
The periods of this curve determine $a$ and $a_D$ in terms 
$u$. One finds
\be
a=\frac{\sqrt{2}\Lambda^2 }{2\pi } \int_{-\Lambda^2}^{\Lambda^2} \frac{dx \sqrt{x-u}} {\sqrt{x^2-\Lambda^4}}\ ,\qquad 
a_D = \frac{\sqrt{2}\Lambda^2}{\pi } \int_{\Lambda^2}^u \frac{dx \sqrt{x-u}} {\sqrt{x^2-\Lambda^4}}\ .
\ee
These integrals  can be expressed in terms of elliptic functions. A compact form is \cite{Klemm:1995wp}
\bea
&& a_D(u) = \frac{i}{4}\Lambda (u^2-1) {}_2F_1(\frac{3}{4},\frac{3}{4},2;1-u^2)\ ,
\nonumber\\
&& a(u) =
\frac{1}{1+i}\Lambda (1-u^2)^{\frac{1}{4}} {}_2F_1(-\frac{1}{4},\frac{3}{4},1;\frac{1}{1-u^2})\ .
\label{eliptica}
\eea
The prepotential ${\cal F}(a)$ can then be obtained from the formula
\be
a_D = \frac{\partial {\cal F} }{\partial a}\ .
\ee
At weak coupling, one finds the expansion \cite{Klemm:1995wp}
\be
\label{swin}
2\pi i {\cal F} = -4 a^2 \ln \frac{4a}{e^{3/2}\Lambda} +\sum_{k=1}^\infty z_k \frac{\Lambda^{4k}}{a^{4k-2}}\ ,
\ee
with 
\be
\label{swins}
z_1= \frac{1}{2^5}\ ,\quad z_2=\frac{5}{2^{14}}\ ,\quad z_3= \frac{3}{2^{18}}\ ,\quad z_4=\frac{1469}{2^{31}}\ ,...
\ee

\subsection{Localization}

We wish to reproduce the formula for the prepotential (\ref{swin}) starting with the exact formula for the partition function
for the theory compactified on a four-sphere of radius $R$, derived by using localization techniques. 
For pure ${\cal N}=2 $ SYM, the one-loop determinant is divergent and needs to be properly
regularized and renormalized. 
An elegant way to obtain the renormalized partition function is by 
adding a hypermultiplet of mass $M$ and then taking a suitable limit $M\to\infty $ \cite{Pestun:2007rz}.
The  ${\cal N}=2 $ SYM with a massive hypermultiplet is the familiar
 ${\cal N}=2^*$ theory, which can be viewed as a flow between ${\cal N}=4$ SYM and pure
 ${\cal N}=2$ SYM.
It is a finite theory, since in the UV regime it flows to ${\cal N}=4$ SYM. 
 For large mass $M$, the theory can be viewed as regularized pure  ${\cal N}=2$ SYM, 
where $M$ represents a UV cutoff.
For $SU(2)$ gauge group, the partition function is given by \cite{Pestun:2007rz}
\be
\label{porta}
Z^{{\cal N}=2^*} = \int_{-\infty}^\infty da \ a^2\ e^{-\frac{16\pi^2}{g^2}\ a^2R^2}
\frac{H^2(2aR)}{H(2aR+MR)H(2aR-MR)}\ \big|Z_{\rm inst}^{{\cal N}=2^*}(a,M)\big|^2\ ,
\ee
\be
H(x)=\prod_{n=1}^\infty \left( 1+\frac{x^2}{n^2}\right)^n e^{-\frac{x^2}{n}}\ ,
\nonumber
\ee
which as expected is a convergent expression. The Nekrasov  instanton function $Z_{\rm inst}^{{\cal N}=2^*}(a,M)$ is computed with equivariant
parameters $\epsilon_{1}=\epsilon_2=1/R$ \cite{Okuda:2010ke}.

The factor $e^{-\frac{x^2}{n}}$ --which renders the infinite product convergent-- is not present
automatically in the one-loop determinant. For the ${\cal N}=2^*$ theory, one is free to add it since
it cancels out between numerator and denominator (modulo a constant). 

The function $H$  can be written in terms of the Barnes G-function, 
$$
H(x)=e^{-(1+\gamma) x^2} G(1+ix) G(1-ix)\ .
$$
For large argument, it has the asymptotic form
\be
\label{asis}
\ln H(x) = -x^2\ln |x| e^{\gamma-\frac{1}{2}}+O(\ln x)\ .
\ee
Thus, for large $M$, the partition function takes the form
\be
Z^{{\cal N}=2^*}  \to e^{2M^2R^2\ln MR} 
\int_{-\infty}^\infty da \ a^2\ e^{-\frac{16\pi^2}{g_R^2}\ a^2R^2}
H^2(2aR)\ \big|Z_{\rm inst}(a)\big|^2\ ,
\ee
where $g_R^2$, defined by,
\be
\label{reno}
\frac{4\pi^2}{g_R^2}\equiv \frac{4\pi^2}{g^2} - 2\ln MRe^{1+\gamma}\ ,
\ee
is kept fixed. 
The coupling $g_R^2$
represents the renormalized coupling  at the scale set by the radius of the four-sphere, and
the factor $e^{2M^2R^2\ln MR} $ reproduces the expected UV divergence of the partition function coming from zero modes of the one-loop determinant.

As usual in asymptotically free theories, $g_R^2$ should  be traded by the dynamical scale of the theory:
\be
\label{limon}
\frac{1}{2}\, \Lambda R \equiv  \lim_{M\to\infty,\ g\to 0} \ MR  \ e^{-\frac{2\pi^2}{g^2}}= e^{-\frac{2\pi^2}{g_R^2}-1-\gamma}\ ,
\ee
which is the only parameter in the problem. Thus
\be
\label{heras}
Z= {\rm const.} \int_{-\infty}^\infty da \ a^2\ e^{8 a^2R^2\ln (\frac{1}{2}\Lambda Re^{1+\gamma})}
H^2(2aR)\ \big|Z_{\rm inst}(a)\big|^2\ .
\ee
The instanton factor of ${\cal N}=2^*$ flows automatically to the instanton factor for pure SYM, once
taken into account the renormalization (\ref{reno}), 
with no extra divergent factor.
For example, for  one-instanton and two-instantons, one has 
$$
Z_{\rm inst}^{{\cal N}=2^*}(a,M) =z_1\ q + z_2 \ q^2+ ... ,
$$
$$
q=e^{2\pi i\tau},\quad \tau =\frac{\theta}{2\pi}+ \frac{4\pi i}{g^2}\ ,
$$
where $z_1$ and $z_2$ are given in (\ref{z1z2}), with $\epsilon=1/R$.
The $\theta $ parameter plays no r\^ole in our discussion so it will be set to zero.
For $M\to\infty $ and $g\to 0$ with $\Lambda $ fixed, we find
\bea
z_1 \  e^{2\pi i\tau } & \to & \frac{\Lambda^4}{2^4}  \frac{1}{ 2\epsilon ^2 \left(a^2+\epsilon ^2\right)}\ ,
\nonumber\\
z_2  \ e^{4 \pi i\tau } & \to & \frac{\Lambda^8 }{2^{8}}\frac{8 a^2+33 \epsilon ^2}{ 4\epsilon ^4 \left(a^2+\epsilon ^2\right) \left(4 a^2+9 \epsilon ^2\right)^2}\ .
\label{inton}
\eea
Thus, this limit just decouples the hypermultiplet, giving rise to the correct finite  instanton coefficients of pure SYM on the four-sphere.

\subsection{Partition function at large $R$ }

Let us return to the computation of the partition function.
For theories with coupling constant, such as ${\cal N}=2^*$, the radius is an independent parameter
that can be sent to infinity at any fixed coupling.
However, for pure $SU(2)$ SYM,  the partition function depends only on one parameter $\Lambda R$. One may explore the theory in the ultraviolet (weak coupling), $\Lambda R\ll 1$, 
or in the  infrared  (strong coupling), where $\Lambda R\to \infty$. 

In the weak coupling limit $\Lambda R\ll 1$,  one can use the above formula (\ref{heras}) to compute the perturbation series to any desired loop order,
just by Taylor expanding $H(2aR)$ in powers of $2aR$. In the process, one discovers that  perturbation series  has a finite radius of convergence \cite{Aniceto:2014hoa}.

Our main interest here is to see if we can find a closed, analytic form in
  the strong coupling regime, $\Lambda R\gg 1$.
Fixing $\Lambda $, this implies looking at the decompactification limit $R\to \infty$. 
We write
\be
\label{patal}
Z= \int da \ e^{-R^2 S(a)}\ ,
\ee
with
\be
R^2 S(a) = - \ln a^2R^2 - 8 a^2R^2\ln (\frac{1}{2}\Lambda Re^{1+\gamma}) -2 \ln H(2aR) 
- \ln Z_{\rm inst}(a)- \ln \bar Z_{\rm inst}(a)
\ee
Since $R$ is large, it is natural to assume that, in this limit, the integral (\ref{patal})
will be dominated
by a saddle point. Assuming that the saddle point lies at a real $aR\gg 1$
--which we will turn to be a self-consistent assumption-- then one can use
 the asymptotic expansion (\ref{asis}) to show that $\ln H $ scales with $R^2$.
Similarly, considering the above one instanton and two instanton terms (\ref{inton}), we find
\be
\ln Z_{\rm inst}(a)\to R^2 \left(\frac{1}{2^{5}}\ \frac{\Lambda^4}{a^2}  +
\frac{5}{2^{14}}\ \frac{\Lambda^8}{a^{4}}+...\right)\ .
\ee
This exactly reproduces the instanton expansion of the prepotential (\ref{swin}), (\ref{swins})
of $SU(2)$ SYM obtained from Seiberg-Witten theory. This is not a surprise, it follows from the universal formula \cite{Nekrasov:2002qd}
\be
\label{nek}
2\pi i {\cal F}_{\rm ins} (a) = \lim_{\epsilon_{1,2}\to 0} \epsilon_1\epsilon_2 \ln Z_{\rm ins}\ ,
\ee
upon making the identification $\epsilon_{1}=\epsilon_2=1/R$. 
Therefore, we find
\be
S(a) \to  8 a^2\ln \frac {4\, a}{e^{\frac{3}{2}}\, \Lambda }
-\frac{\Lambda^4}{2^4 a^2}  -
\frac{5}{2^{13} }\ \frac{\Lambda^8}{a^{4}}+...
\ee
This is nothing but twice the prepotential (\ref{swin}) of $SU(2)$ ${\cal N}=2$ supersymmetric Yang-Mills theory, including the one-loop term.
Note that the one-loop term has combined with the Gaussian term coming from the curvature coupling of the scalar field of the vector multiplet.
Thus 
\be
\label{apest}
\lim_{R\to \infty } \frac{1}{R^2}\ \ln Z_{\rm Pestun} ({\mathbb S}^4) =
 2\pi i \left({\cal F}(a)-\bar {\cal F}(a)\right)\ .
\ee
It is natural to conjecture that this formula extends to any  ${\cal N}=2$ theory with arbitrary gauge group and matter content, in a regime of coupling where  saddle points
at large $a_i$ exist.

Without the instanton terms, the saddle-point calculation for the  ${\cal N}=2$ $SU(2)$ SYM theory was carried out in \cite{Russo:2012ay}, finding that there
was indeed a non-trivial saddle point dominating the integral. 
The calculation in \cite{Russo:2012ay} was made in the context of a toy model, in order  to motivate  large $N$ physics.
It was a toy model because instantons cannot be ignored for $SU(2)$, 
 as we will shortly confirm. 

We look for a saddle point on the real line for $a$, so $a=\bar a$. 
The saddle-point equations are now 
\be
\frac{\partial S}{\partial a}=0 \ \longrightarrow \ \frac{\partial {\cal F}}{\partial a}=0\ .
\ee
Hence
\be
a_D = 0\ .
\ee
Strikingly, as long as there is a saddle point at large $aR$ dominating the integral, the exact determination of the partition function amounts to computing
the prepotential at the point where the dual magnetic variable vanishes and a monopole becomes massless.
Since $a_D$ is a period integral,
$a_D=0$ represents a singularity of the curve.
For the $SU(2)$ curve (\ref{curr}), this is the singularity located at  $u=\Lambda ^2$. At this point
\be
a\to a^*=\frac{2\Lambda}{\pi}\ ,\qquad a_D\to 0\ .
\ee
In particular, this confirms that, at the saddle point, instanton effects are of order 1,
since $a\sim \Lambda $. Therefore they cannot be neglected.
It also confirms that the saddle point occurs at $aR\gg 1$, provided $\Lambda R\gg 1$.

The prepotential can be computed from the Matone relation \cite{Matone:1995rx}:
\be
u=2\pi i \left({\cal F}(a) -\frac{1}{2}a\partial_a {\cal F}(a) \right)\ ,
\ee
which, at the saddle point, gives
\be
\label{koi}
2\pi i{\cal F}(2\Lambda/\pi) =  \Lambda ^2\ .
\ee
Thus
\be
\label{sodas}
\lim_{R\to \infty} \frac{1}{R^2}\ \ln Z_{\rm Pestun}({\mathbb S}^4)  = 2\Lambda^2 \ .
\ee
This may be compared with the result of \cite{Russo:2012ay} for the toy model without instantons, $\ln Z\sim  e^{-2\gamma}\Lambda^2 R^2$.

\begin{figure}[h!]
\centering
\includegraphics[width=0.6\textwidth]{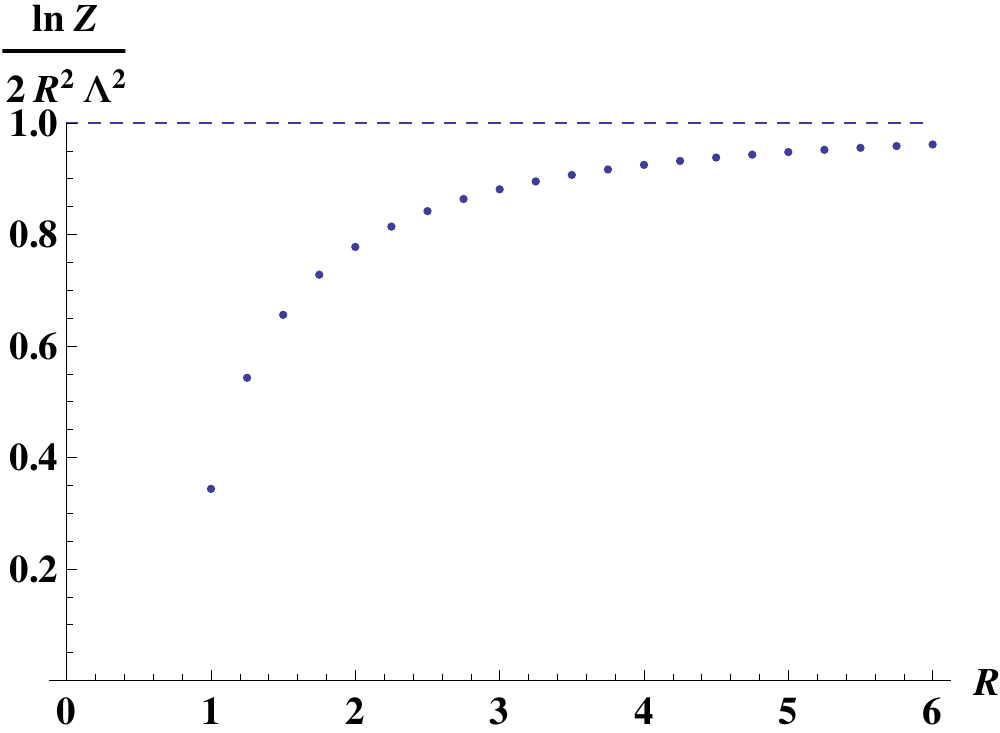}
\caption{Ratio between the $\ln Z$ computed numerically using the exact Seiberg-Witten solution and the saddle-point estimate $\ln Z\big|_{{\rm saddle}} \approx 2\Lambda^2 R^2$ for different values of $R$.}
\label{saddleSYM}
\end{figure}

In order to justify   the saddle-point  approximation, we need to compute
 the second derivative of the action. This gives
\be
\frac{\partial^2 S}{\partial a^2}= - 4\pi i \frac{\partial a_D }{\partial a} =-4\pi i\tau(a)\ .
\ee
Note that $-4\pi i\tau(a)$ is real and $>0$. From the behavior near the singularity,  
\be
\label{loi}
a_D\approx \frac{i}{2\Lambda }(u-\Lambda^2)\ ,\qquad a\approx \frac{2\Lambda }{\pi}-\frac{1}{4\pi \Lambda} (u-\Lambda^2)\ln (\frac{u}{\Lambda^2}-1)\ ,
\ee
we find
\be
R^2 \frac{\partial^2 S}{\partial a^2}\approx  \frac{ 8\pi^2 R^2}{|\ln (\frac{u}{\Lambda^2}-1) | }\to 0\ .
\ee
This is consequence of the familiar fact that the electric coupling diverges at the point where the monopole is massless.
Although the second derivative of the action vanishes, all higher derivatives, however, diverge at this point, showing that this point has more the structure of a  cusp than a Gaussian shape.
Nevertheless, because of the  sharp peak at $a^*$,  the saddle point still captures the leading behavior in $\ln Z$, despite $R^2 S''= 0$ at the singularity.  This is shown in fig. \ref{saddleSYM},
which shows the ratio between  the partition function 
computed numerically at large $R$, using the exact Seiberg-Witten solution (\ref{eliptica}), and the saddle-point result (\ref{sodas}).
We have used that, at large $R$,
\be\label{jare}
Z= \int da \ e^{4\pi i R^2 {\cal F}(a)}= \int du\ \partial_u a\  e^{4\pi i R^2 {\cal F}(u)}\ ,
\ee
where ${\cal F}(u)$ is obtained by integrating $a_D(u) \partial_u a(u)$ using (\ref{eliptica})
and (\ref{koi}). 
 In (\ref{jare}), 
the contour  in the  integral over $u$  has been chosen from $u=\Lambda^2$  to infinity on the real axes. The integral approaches the   same value for any contour passing near  $u=\Lambda^2 $, because at large $R$ the integral is dominated by the $u=\Lambda^2$ region.
In the original integral, this choice of contour corresponds to $a$ going from $2\Lambda/\pi$ to infinity, where we have
chosen the branch where $a$ is real (then the prepotential is purely imaginary and the action is real). Other branches around $\Lambda ^2$ have imaginary components for $a$, as can be seen
from the monodromy $a_D\to a_D\ ,\ a\to a-a_D$ arising as $u$  circles 
the singularity at $u=\Lambda^2$ (cf. (\ref{loi})).
While the original partition function (\ref{patal}) involves an integration from $a=0$ to $\infty$,
in the Seiberg-Witten quantum solution there is no contour in the $u$-plane where $a(u)$ is real and $0<a<2\Lambda/\pi $.  It would be interesting to compute the complete partition function 
 (\ref{patal})  numerically at finite $R$. This requires knowing the instanton factor in a closed form. 

\section{$SU(2)$ SQCD with  massive fundamental and antifundamental hypermultiplets}

\subsection{The partition function}

Consider now ${\cal N}=2$ supersymmetric Yang-Mills theory coupled to $N_f=2$ massive
matter, namely
a fundamental and an antifundamental hypermultiplet of mass $M$.
This theory is asymptotically free.
Localization now leads to the following partition function\footnote{
In \cite{Russo:2013kea,Russo:2013sba} $N_f$ represents $N_f$  pairs of fundamental and antifundamental hypermultiplets, so our  model here corresponds to $N_f=1$ and $N=2$ in the notation of
\cite{Russo:2013kea,Russo:2013sba}.}
\be
\label{sqqq}
Z^{\rm SQCD} ({\mathbb S}^4) = {\rm const.} \int_{-\infty}^\infty da \ a^2\ 
e^{4 a^2R^2\ln \Lambda Re^{1+\gamma}}
\frac{ H^2(2aR)}{H^2(aR+\frac{MR}{\sqrt{2}})H^2(aR-\frac{MR}{\sqrt{2}})}
\ \big|Z_{\rm inst}(a)\big|^2\ .
\ee
This partition function is obtained after renormalization procedure.
Like in the case of pure SYM theory, this can be carried out 
by starting from a finite theory, the theory obtained by adding two extra (fundamental and antifundamental) hypermultiplets of mass $M_0$ and using the flow from 
the superconformal ${\cal N}_f=4$ theory  to ${\cal N}_f=2$. 
 This approach was followed in  \cite{Russo:2013sba}.
In the UV, the resulting theory flows to the  ${\cal N}=2$ superconformal SYM;
therefore it is a finite theory with convergent partition function. The idea is then to take the $M_0\to \infty $
limit to decouple the two extra hypermultiplets by following similar steps as we did for pure SYM.
The limit   leads to the partition function (\ref{sqqq}), with the identification
\be
\lim_{M_0\to\infty, \ g_0\to 0}M_0 e^{- \frac{4\pi^2}{g_0^2}} = \Lambda ={\rm fixed}\ ,
\ee
where $g_0$ is the coupling of the original theory.

We now take the decompactification limit. This implies looking into the infrared regime,
where $\Lambda R\gg 1$. We assume again that in this limit the integral is dominated
by a saddle point at some $aR\gg 1$.
Using the asymptotic formula (\ref{asis}) for $H$ and the formula (\ref{nek}), we now find
\be
\label{sarr}
\lim_{R\gg 1 }   Z^{\rm SQCD} ({\mathbb S}^4) =\int da \ e^{-R^2 S(a,M)}\ ,
\ee
with
\bea
S(a,M) &=& 8 a^2 \ln \frac{2e^{\frac{1}{4}}\, a}{\Lambda} -2(a+\frac{M}{\sqrt{2}})^2\ln
\frac{|a+\frac{M}{\sqrt{2}}|}{\Lambda} -2(a - \frac{M}{\sqrt{2}})^2\ln
\frac{|a-\frac{M}{\sqrt{2}}|}{\Lambda} 
\nonumber
\\
&-& 2\pi i {\cal F}_{\rm ins} + 2\pi i \bar {\cal F}_{\rm ins} \ .
\eea
We recognize the one-loop contribution to the prepotential, which, combined  with the instanton contributions, gives the full prepotential of the theory \cite{ Ohta:1996fr,D'Hoker:1996nv}.
The singularity at $a=\pm M/\sqrt{2}$ represents the point in the moduli space where
the  hypermultiplet becomes massless.

Note that, once again, the $\ln \Lambda $ piece originating from the curvature coupling of the scalar field of the vector multiplet has combined with the terms from the one-loop determinant to produce the
correct one-loop terms of the prepotential with the dynamical scale $\Lambda $ included.
Thus 
\be
\label{zqcd}
\lim_{R\to \infty}   \frac{1}{R^2} \ln Z^{\rm SQCD} ({\mathbb S}^4) =   2\pi i {\cal F}(a^*) - 2\pi i \bar {\cal F}(a^*) \ ,
\ee
provided $a^*R\gg1 $. The limit is taken with $M, \ \Lambda $ fixed.
To complete the derivation, we need to find the saddle point, compute the prepotential at the saddle point and show that the approximation is justified.

\subsection{SQCD toy model without instantons}

The basic physical mechanisms underlying  the large $N$ phase transitions of \cite{Russo:2013kea,Russo:2013sba} can be illustrated in the $SU(2)$ SQCD model by 
ignoring the instanton terms. It is a toy model  because, as shown below,  instantons cannot be cannot be neglected in any regime of the coupling. The model, however, contains the essential ingredients of the large $N$ $SU(N)$ models of \cite{Russo:2013kea,Russo:2013sba} that exhibit phase transitions.

The saddle point  corresponds to the minimum of the action:
\be
S_0(a,M) =8 a^2 \ln \frac{2e^{\frac{1}{4}} a}{\Lambda }-2 (a-m)^2 \ln \frac{|a-m|}{\Lambda
   }-2 (a+m)^2 \ln \frac{|a+m|}{\Lambda }\ ,
\ee
with $m\equiv M/\sqrt{2}$.
The saddle-point equation is then given by
\be
\label{abel}
0=a+4 a \ln \frac{2e^{\frac{1}{4}} a}{\Lambda }-(a-m) \ln \frac{|a-m|}{\Lambda
   }-(a+m) \ln \frac{|a+m|}{\Lambda }\ .
\ee
This is a transcendental equation which can be solved  analytically in different regimes.
Let us call $a^*$ the value of $a$ at the saddle point.
Then, as usual, the partition function is given by
\be
\label{sado}
\ln Z \to -R^2 S_0(a^*)\ .
\ee
One can numerically verify  that, as expected, the saddle-point formula (\ref{sado})
reproduces the complete integral over $a$ in $Z$ with arbitrary accuracy for sufficiently large $R$.\footnote{The convergence is much faster if one includes the
quadratic fluctuations and uses  $Z= 2\sqrt{2\pi}/(R\sqrt{S''(a^*)}) \ e^{-R^2 S(a^*)} $.}

In the weak coupling regime, $\Lambda \ll m$, and the minimum is at small values of $a$, 
Expanding (\ref{abel}) in powers of $a$, we find
the solution
\be
a^*=\frac{1}{2} \sqrt{m  \Lambda } \left( 1- \frac{1}{48} \frac{ \Lambda}{m}-
\frac{11}{23040} \left(\frac{ \Lambda}{m}\right)^2+...\right)\ ,
\ee
and
\be
F=- \ln Z = R^2m^2 \left( 4 \ln \frac{ \Lambda }{m}-  \frac{ \Lambda}{  m}+\frac{1}{48}\frac{ \Lambda ^2}{m^2}+\frac{1}{5760}\frac{ \Lambda
   ^3}{ m^3}+\frac{1}{5760}\frac{ \Lambda
   ^4}{ m^4}+... \right)\ .
\ee
This can be recognized as an OPE expansion in terms of the dynamical scale $\Lambda$.
\footnote{The emergence of an OPE expansion at weak coupling was  noticed in \cite{Russo:2013qaa,Russo:2013kea}
for the  $SU(N)$ models.}

As $\Lambda/m $ is increased, the minimum $a^*$ increases until it hits the singularity at $a=m$
where a component of the elementary hypermultiplet becomes massless. 
This occurs at
\be
\Lambda_c=2 e \, m\ .
\ee
This is the analog of the critical point in the large $N$ phase transitions of  \cite{Russo:2013kea,Russo:2013sba}.
Near  $\Lambda_c$, 
\be
\Lambda-\Lambda_c \approx e  \  (m-a^*)\ln (1-\frac{a^*}{m})\ .
\ee

As $\Lambda $ is further increased and becomes greater than $\Lambda_c$, $a$ crosses the massless singularity and keeps increasing.
For large $\Lambda/m $, $a^*$ is large and we find the behavior
\be
a^*\approx \frac{\Lambda}{4e}\Big(1+O\big(\frac{ m^2}{\Lambda^2}\big)\Big)\ ,
\quad F\approx -\frac{1}{8e^2}\, \Lambda^2 R^2  
\Big(1+O\big(  \frac{m^2}{\Lambda^2}\big)\Big)\ .
\ee

We can now see that neglecting instantons cannot be justified  in any of 
the above three regimes. By looking at the first few terms in the instanton expansions (see e.g.  \cite{Ohta:1996fr}), one finds that instanton effects are small provided:
\smallskip

\noindent a) $a\gg \sqrt{\Lambda M}$, in the weak coupling regime $\Lambda\ll M$.
\smallskip

\noindent  b) 
$a\gg \Lambda $, near $\Lambda_c$ or  in the strong coupling regime $\Lambda\gg M$.

\smallskip

\noindent Comparing with the values of the saddle points $a^*$ given above, we see that in no case
instanton contributions can be neglected.

In the following section we will see how instantons affect this picture.

\subsection{Exact results via Seiberg-Witten}

The SW curve for ${\cal N}=2$ $SU(2)$ SYM with two  flavors of equal mass is
\be
\label{masw}
y^2 = \left(x^2- \frac{1} {64} \Lambda^4\right)  (x-u)+ \frac{1}{4} M^2\Lambda^2 \ x - \frac{1}{32}M^2\Lambda^4\ .
\ee
In this case, $a$ and $a_D$ are defined as period integrals of the meromorphic one-form
\be
\lambda = -\frac{\sqrt{2}}{4\pi } \frac{ y\, dx}{x^2-\frac{\Lambda^4}{64}}\ .
\ee
By a shift $x\to x+u/3$, we can write the curve (\ref{masw})  in the Weierstrass form
\be
y^2 = (x-e_1 ) (x-e_2 ) (x-e_3 )  \ ,
\ee
with
\bea
e_1 &=& \frac{u}{6} -\frac{\Lambda^2}{16}+\frac{1}{2}\sqrt{ u+\frac{\Lambda^2}{8}+\Lambda M}\sqrt{ u+\frac{\Lambda^2}{8}-\Lambda M}\ ,
\nonumber\\
e_2 &=& -\frac{u}{3} +\frac{\Lambda^2}{8}\ ,
\nonumber\\
e_3 &=& \frac{u}{6} -\frac{\Lambda^2}{16}-\frac{1}{2}\sqrt{ u+\frac{\Lambda^2}{8}+\Lambda M}\sqrt{ u+\frac{\Lambda^2}{8}-\Lambda M}\ .
\eea
It has singularities at the zeroes of the discriminant
\be
\Delta = \frac{1}{2^{16}} \Lambda ^4 \left( \Lambda ^2+8M^2-8u\right)^2 \left(\left( \Lambda
   ^2+8u\right)^2-64 \Lambda ^2 M^2\right)\ ,
\ee
i.e. at
\be\label{singo}
u_1=-M\Lambda -\frac{\Lambda^2}{8}\ ,\quad
u_2=M\Lambda -\frac{\Lambda^2}{8}\ ,\quad
u_3=M^2+\frac{\Lambda^2}{8}\ .
\ee
The periods $a$ and $a_D$ for this curve were explicitly computed in \cite{Bilal:1997st}.
$a_D$ is defined as an integral over the cycle $\gamma_2$ surrounding  $e_1$ and $e_2$,
whereas $a$ on the cycle $\gamma_1$ surrounding  $e_2$ and $e_3$.
The cycle  $\gamma_1$ picks also a pole of the  one-form $\lambda$
whose residue is $M/\sqrt{2}$.

One of the salient aspects of this theory is the occurrence of an Argyres-Douglas \cite{Argyres:1995jj}
superconformal fixed point \cite{Argyres:1995xn,Eguchi:1996vu}. 
This arises when some zeroes of $\Delta $ coincide. Then, at the singularity, $e_1$, $e_2$ and $e_3$ get together and the Riemann surface develops a cusp.
From  (\ref{singo}),
we see that this occurs at
\be
2M=\Lambda \ .
\ee
An important question is whether there is any manifestation of the existence of this fixed point in the partition function.
We have argued that in the large $R$ limit the
 partition function on the four-sphere (\ref{sqqq}) can be dominated by a saddle point if
the action has a minimum at $a^*R\gg 1$. In such a case, the partition function 
can be read from the prepotential evaluated at the saddle point, as prescribed  in (\ref{zqcd}).  

\begin{figure}[h!]
\centering
\begin{tabular}{cc}
\includegraphics[width=0.5\textwidth]{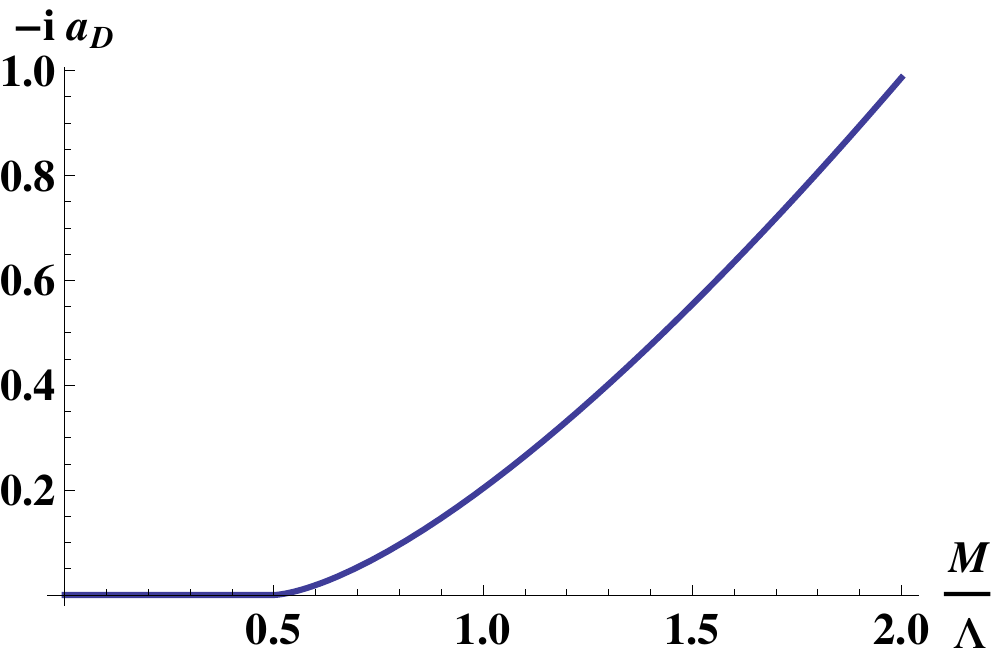}
&
\includegraphics[width=0.55\textwidth]{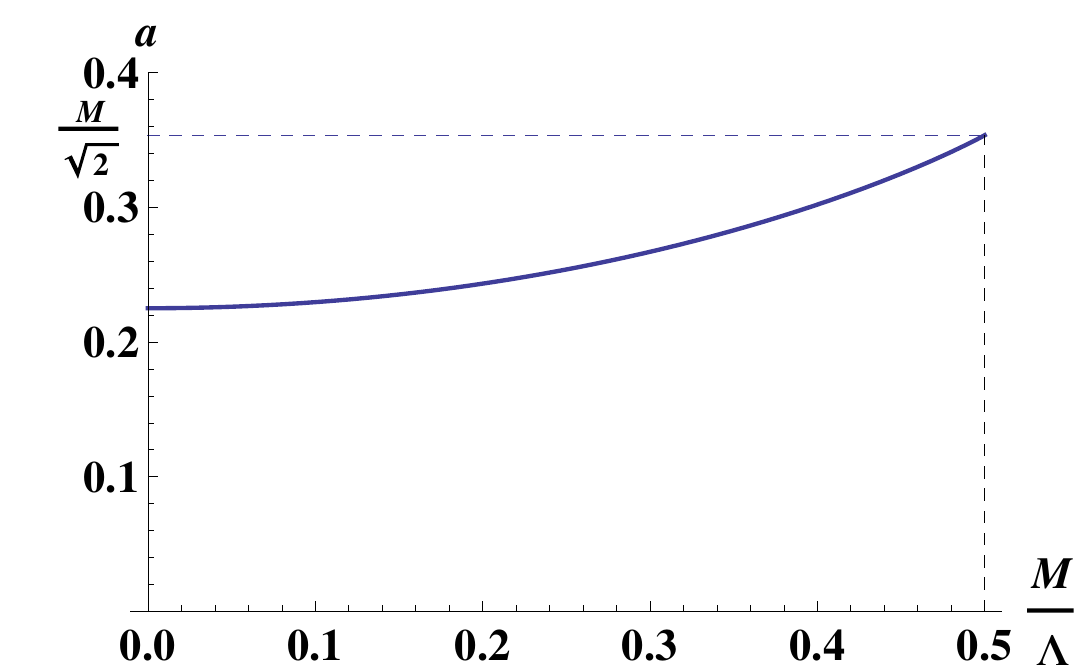}
\\
(a)&(b)
\end{tabular}
\caption{a) $-i a_D$ as a function of $M/\Lambda $ at the singularity $u=M^2+\frac{1}{8}\Lambda^2$.
It vanishes identically for $M< \Lambda/2$, showing that it corresponds to a saddle point
in the partition function. b)  $a$ as a function of $M/\Lambda $ at the same singularity.}
\label{aDaa}
\end{figure}

The saddle-point equation is
\be
\frac{\partial S(a,M)}{\partial a}=0\ \longrightarrow  \ a_D=\frac{\partial {\cal F}}{\partial a} = 0\ .
\ee
The behavior of $a_D$ was examined in detail in \cite{Bilal:1997st}, and can be  understood
by looking at the above expressions for $e_1,\ e_2,\ e_3$.
The equation $a_D=0$ requires that $e_1\to e_2$. 
This is the singularity with
\be
u_3= M^2+\frac{1}{8}\Lambda^2 \ .
\ee
More precisely, this gives $e_1=e_2$ provided $M<\Lambda/2$.
When $M>\Lambda/2$, then one has $e_2=e_3$ and $a_D\neq 0$.
To verify this, we  examine the exact formula for $a_D$ in terms of elliptic integrals
(eq. (2.27) in \cite{Bilal:1997st}). Figure \ref{aDaa}a shows a  plot of  $-ia_D$ as a function of $\Lambda/M$,
which confirms that $a_D$ is nowhere vanishing when 
$M>\Lambda/2$. 

\begin{figure}[h!]
\centering
\includegraphics[width=0.6\textwidth]{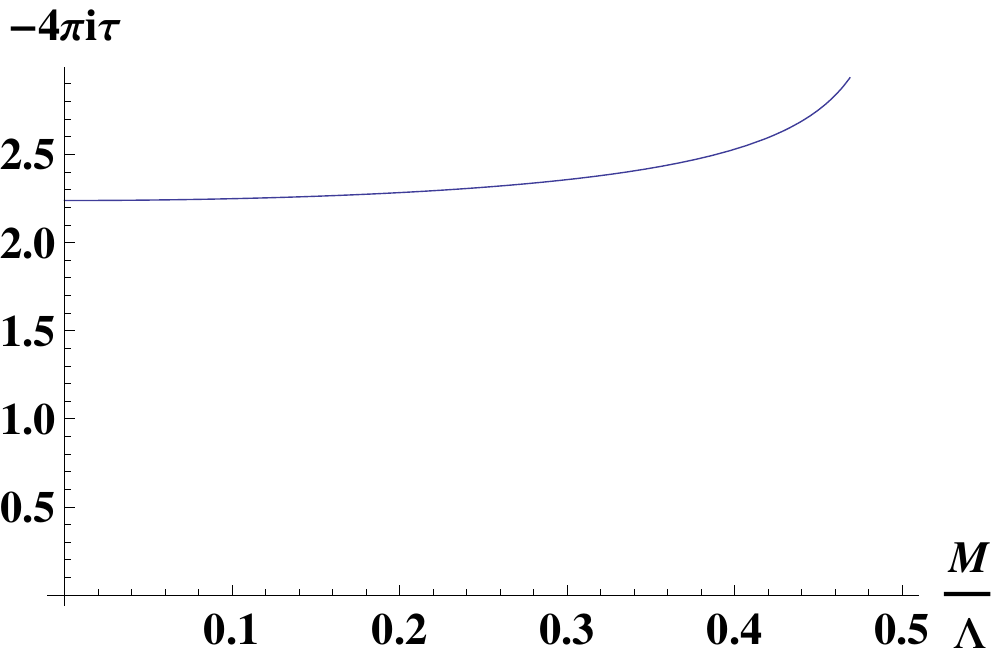}
\caption{$-4\pi i\tau(a,M)$ as a function of $M/\Lambda$ on the saddle point at $e_1=e_2$,
$u=M^2+\frac{1}{8}\Lambda^2$, $M<\Lambda/2$.}
\label{tauta}
\end{figure}

Thus a saddle point exists only when $M<\Lambda/2$. To justify the saddle-point approximation, we must compute the second derivative of the action. This gives
\be
R^2\frac{\partial^2 S(a,M)}{\partial^2 a}= -R^2 4\pi i\tau(a,M) >0 \ .
\ee
The coupling $-4\pi i \tau(a,M)$ at the $u_3$ singularity is shown in figure \ref{tauta} as a function of $M/\Lambda $. We see that the second derivative of $S$ is positive and $O(1)$ in the whole interval $0<M<\Lambda/2$.
Therefore, $R^2 S''\to \infty $ in the infinite $R$ limit and the saddle-point approximation becomes exact.

The behavior of the saddle-point $a$ at the singularity, $e_1=e_2$ is
 shown in fig. \ref{aDaa}b. We have used the exact expressions in terms of elliptic integrals given in  \cite{Bilal:1997st} (this picks the specific branch where $a(u_3)$ is real).
Importantly,
\be
\lim_{M\to \frac{\Lambda}{2}} a = \frac{M}{\sqrt{2}}\ .
\ee
This is a consequence of the fact that at this point $e_2\to e_3$ and the period integral over $\gamma_1$ vanishes. When $M\to \Lambda/2$, the integral defining the partition function is dominated by a saddle point located precisely at the point where a component of the hypermultiplet becomes massless. From fig.  \ref{aDaa}b
we see that the value of $a$ increases from a non-zero value $a=\frac{1}{\sqrt{2}\pi}$ at $M=0$, until it hits the singularity at $M\to \Lambda/2$.
As long as $M<\Lambda/2$, the free energy will be given by
$F= -R^2 {\rm Re} \big(4\pi i {\cal F}(a^*)\big)$.
On the other hand, when $M>\Lambda/2$ computing the free energy requires
an integration over the full domain, as there seems to be no saddle point dominating the integral.
This gives evidence of non-analytic behavior of the free energy as a function of the coupling
$\Lambda/M$ in crossing the point
 $M=\Lambda/2$, and therefore a phase transition.
At the critical point, the theory is described by an interacting superconformal theory, whose  spectrum of scaling dimensions was discussed in \cite{Argyres:1995xn}.

The free energy is thus completely determined in the strong coupling phase
$M<\Lambda/2$ in terms of the prepotential as a function of $M/\Lambda$, obtained by sitting
on the $u=u_3$ singularity.
To compute the order of the phase transition, one would need  the expression for the free energy in the weak-coupling phase, which we do not know.\footnote{In the large $N$ SQCD model,
the analogous phase transition is third order \cite{Russo:2013kea}.}
 From
the free energy in the weak-coupling phase one can also compute the weak-coupling
OPE expansion for the full model including instanton contributions. In particular, this would be interesting in order to have a better understanding of a long-standing question in QCD, 
concerning the precise manner  by which instanton and non-instanton power-like corrections 
contribute, and how they can be distinguished.\footnote{We thank K. Zarembo for this remark.} 

It is interesting to see how the theory behaves for complex mass parameter.
In this case, the partition function still has a saddle-point at $u=u_3$ for ${\rm Re}(M)<\Lambda/2$, where $e_1\to e_2$.
However, the simultaneous condition $e_1\to e_2$ and $e_2\to e_3$ cannot be satisfied in this case and, as a result, there is no phase transition for any  value of $\Lambda$. The same applies to  the large $N$ $SU(N)$ models \cite{Russo:2013kea}. The saddle-points occur at
real expectation values of the scalar field of the vector multiplet. Therefore, for any real value of $\Lambda $,  they cannot hit the massless singularity, which for complex mass is  located at  complex values of $a$.

In conclusion,  the SQCD $SU(2)$ theory with two flavors seems to have
a  phase transition of a similar nature as the large $N$ phase transition found in
SQCD with $N_f<2N$ flavors discussed in \cite{Russo:2013kea,Russo:2013sba}.
However, note that the picture is very different from what was found in the toy model of  section 3.2. This was expected, since, as shown, instantons are important in the
whole range of couplings.
Nonetheless, just as in the transitions of \cite{Russo:2013kea,Russo:2013sba},
here the phase transition occurs because,  at some critical coupling,
the saddle-point $a$  hits the  singularity where the electric hypermultiplet becomes massless.
We now also see that the critical point of these transitions is precisely the Argyres-Douglas superconformal point of the theory discovered in \cite{Argyres:1995xn}.

\section{${\cal N}=2^*$ $SU(2)$ SYM}

The exact partition function for ${\cal N}=2^*$ $SU(2)$ Super Yang-Mills theory on ${\mathbb S}^4$ is given by the formula (\ref{porta}) derived by Pestun \cite{Pestun:2007rz}.
Different properties of this theory have been discussed in  \cite{Pestun:2007rz} and in subsequent works.  In particular, the vacuum expectation value of a supersymmetric 't Hooft loop operator carrying magnetic charge is computed in \cite{Gomis:2011pf}.  Extending previous results in flat space \cite{Minahan:1997if}, 
Billo {\it et al.} \cite{Billo:2013fi} find that the prepotential terms satisfy a modular anomaly equation, which in turn generates a recursion relation for the coefficients of the expansion in inverse powers of $a$.
The perturbation series for the partition function was studied in \cite{Russo:2012kj,Aniceto:2014hoa}, where it was found that it has an $n!$ large order behavior
associated with Borel singularities originating from zero modes of the one-loop determinant
that occur in the complex $a$-plane.

Our purpose here is to  connect this partition function with the Seiberg-Witten solution
and look for possible phase transitions.
The partition function depends on two independent parameters, $MR$ and the coupling $g$.
We now take the decompactification limit at fixed coupling $g$.
From the asymptotic expansion (\ref{asis}) for $H$, and using Nekrasov formula (\ref{nek}), we find
\be
\lim_{R\to \infty}  \frac{1}{R^2} \ln Z^{{\cal N}=2^*}({\mathbb S}^4) =  - S(a,M)\ ,
\ee
with
\bea
S(a,M) &=& \frac{16\pi^2}{g^2}\ a^2+ 8 a^2 \ln (2|a|R) -(2a+M)^2\ln
|2aR+MR| 
\nonumber
\\
&-& (2a - M)^2\ln|2aR-MR|
-2\pi i {\cal F}_{\rm ins} + 2\pi i \bar {\cal F}_{\rm ins} \ .
\label{azion}
\eea
One  recognizes the classical, one-loop and instanton contributions to the prepotential
(see \cite{Minahan:1997if}).
Thus
\be
S(a,M)=-2\pi i {\cal F} + 2\pi i \bar {\cal F}\ .
\ee

Let us  take the $\epsilon\to 0$ limit in (\ref{nek}) explicitly, by starting with the general
instanton partition function $Z_{\rm inst}^{{\cal N}=2^*}$ on the sphere. 
The one- and two-instanton terms are computed in appendix A.
Using the  expressions for $z_{k=1}$ and $z_{k=2}$ given in (\ref{z1z2}), we obtain
\bea
2\pi i {\cal F}_{\rm ins} (a) &=& \lim_{\epsilon_{1,2}\to 0} \epsilon_1\epsilon_2 \ln Z_{\rm ins}
\nonumber\\
&=& M^2 \left(\Big( \frac{ M^2 }{2 a^2}-2\Big)\ q + \left(\frac{5 M^6}{64 a^6}-\frac{3 M^4}{4 a^4}+\frac{3 M^2}{2 a^2}-3 \right)\  q^2 + ...\right)
\label{variat}
\eea
We can recognize the coefficients $1/2$ and $5/64$ of the instanton expansion (\ref{swin}), (\ref{swins}) in pure SYM,
here appearing as the leading term at large $M$ (which are the only terms that survive upon taking the limit (\ref{limon}).

In the large $N$ limit, the ${\cal N}=2^*$ $SU(N)$ supersymmetric Yang-Mills theory exhibits an infinite number of phase transitions \cite{Russo:2013qaa,Russo:2013kea,Chen:2014vka,Zarembo:2014ooa}. 
In this limit, instantons are exponentially suppressed, so the dynamics of the phase transitions
is fully governed by the classical and one-loop   terms.
To exemplify this dynamics, in appendix B we present a toy ${\cal N}=2^*$ $SU(2)$  model
ignoring instanton contributions. It shows a behavior qualitatively similar to the 
toy  ${\cal N}=2$ $SU(2)$  SQCD described in previous section, with OPE series at weak coupling and a non-analitic behavior at the massless singularity.

The question is whether this picture survives instanton corrections.
To address this question, we  now consider the exact computation of partition function
using Seiberg-Witten description of the model.
As in previous examples, we look for possible saddle points.
The saddle-point equation is
\be
a_D= \frac{\partial {\cal F}}{\partial a}=0\ .
\ee
Denoting this saddle point as $a^*$, we would then have,
\be
\ln Z\sim  2\pi i R^2(  {\cal F}(a^*) -  \bar {\cal F}(a^*) )\equiv R^2 f(MR,g)\ .
\ee
A possible phase transition will occur if $a^*$ reaches $2M$ at some finite coupling $g_{\rm cr}$.
Then possible discontinuities in derivatives of $ f(MR,g)$ with respect to the coupling
at $g_{\rm cr}$ will
dictate the order of the phase transition.

For  the ${\cal N}=2^*$ theory, the Seiberg-Witten curve is \cite{Seiberg:1994aj}
\be
\label{maga}
y^2 = \left(x- e_1 \tilde u -\frac{1}{4} e_1^2 M^2\right) \left(x- e_2 \tilde u -\frac{1}{4} e_2^2 M^2\right) \left(x- e_3 \tilde u -\frac{1}{4} e_3^2 M^2\right)\ .
\ee
where the $e_i$ are the following combinations of  Jacobi $\theta $ functions,
\be
e_1-e_2=\theta^4_3(0,\tau)\ ,\qquad e_3-e_2=\theta^4_2(0,\tau)\ ,\qquad e_1-e_3=\theta^4_4(0,\tau)\ ,
\ee
satisfying $e_1+e_2+e_3=0$, and $\tilde u$ is given by \cite{Dorey:1996ez}
\be
\tilde u= u-\frac{M^2}{12}- M^2\sum_{n=1}^\infty \alpha_n q^n \ .
\ee
Here $u=\frac{1}{2}\langle {\rm tr}\Phi^2 \rangle =a^2+...$.
The numerical values of $\alpha_n$ will not be important for our arguments.
Various aspects of this theory have been extensively studied in the literature (see \cite{D'Hoker:1999ft} and references therein).
A study of quantum critical points in general ${\cal N}=2^*$  $SU(N)$  theories
is in  \cite{Donagi:1995cf}. In particular, 
for a gauge group $SU(3)$, Donagi and Witten find a set of eight 
 Argyres-Douglas critical points, which transform  under the action of $SL(2,{\mathbb Z})$.
It would be interesting to understand these different phases in terms of
the free energy computed by localization.\footnote{On the other hand, the $SU(3)$ {\it pure} SYM has two critical points \cite{Argyres:1995jj}. 
This theory is described in terms of a (genus 2) hyperelliptic Riemann surface 
with six branch points. Minimizing the prepotential requires that the
dual variables $a_{D}^1, a_D^2$ vanish.  Under  this condition 
the $\beta_{1,2} $ cycles shrink. This is, however,  a different condition than the one
leading to the $Z_3$ conformal fixed points of \cite{Argyres:1995jj}.
Presumably the critical points are reached only in a singular limit.
}

Returning to the curve (\ref{maga}), singularities are at
\be
\tilde u_i = \frac{1}{4} e_i M^2 \ ,\qquad i=1,2,3\ .
\ee
The weak coupling expansions for the $e_i$ are  $e_1=2/3+O(q)$, $e_{2,3}=-1/3+O(q^{1/2})$.
Then, at weak coupling, near the singularity $\tilde u_1$,
\be
u\approx \tilde u_1+\frac{M^2}{12} \approx  \frac{1}{4} M^2 \ .
\ee
i.e. $a\approx \pm M/2$, corresponding to the point where a component of the hypermultiplet become massless.
The behavior  of the effective coupling $\tau(a)$ near this singularity was studied  in \cite{Minahan:1997if}. It has the expected classical and one-loop term plus instanton
corrections given in terms of the Dedekind $\eta $ function.
In the Donagi-Witten approach \cite{Donagi:1995cf}, the singular point corresponds to a degenerating limit of a genus 2 Riemann surface.

Defining the branch points as 
$x_i=e_i \tilde u +\frac{1}{4} e_i^2 M^2 $, then
$a$ is defined as a period integral over the cycle  $\gamma_1$ that loops around $x_2,\  x_3$,
and   $a_D$ with $\gamma_2$ that loops around $x_1$ and $x_2$.\footnote{Choosing another combination such as
$x_1$ and $x_3$ shifts $a_D$ by an integer multiplying $a$. This can be removed by an integer shift of $\tau $, since $a_D$ obeys the asymptotic condition $a_D \approx 2a\tau $.}

The saddle-point solution occurs at the value of $u$ where $a_D$  vanishes, which
in turn is the singularity at $\tilde u=\tilde u_3$, producing the shrinking of the $\gamma_2$ cycle, $x_1\to x_2$. Therefore, the solution of the saddle-point equation is
\be
u= \frac{1}{4} e_3 M^2  +\frac{M^2}{12}+ M^2\sum_{n=1}^\infty \alpha_n q^n\ .
\ee
Since $a$ is a function of $u$, this defines $a(\tau)$ on the saddle point.
Our aim is to see if there is a coupling $\tau_c$ such that $a(\tau)$ meets the massless hypermultiplet singularity, i.e. $a(\tau_c)= \pm M/2$.
This requires that at this value of the coupling, $x_2\to x_3$, i.e. that $\tilde u=\tilde u_1$.
However, both conditions together require $\tilde u_1=\tilde u_3$, which  is impossible, since
\be
\tilde u_1-\tilde u_3 =  \frac{1}{4} M^2 \ \theta^4_4(0,\tau)\neq 0\ ,
\ee
as $ \theta^4_4(0,\tau)$  is nowhere vanishing in the upper half complex $\tau $ plane.
Therefore  there cannot be a phase transition in the $SU(2)$ model.
The phase transition appearing in the toy model of appendix B 
vanishes away (or moves to $g\to \infty$) when instanton contributions are incorporated.
This is perhaps expected, given that there is no
 Argyres-Douglas superconformal fixed point in ${\cal N}=2^*$ $SU(2)$ theory at finite coupling, and we have argued
that there is a correspondence between conformal fixed points and quantum critical points in phase transitions associated with
massless resonances.

\subsection*{Acknowledgements}
 
We would like to thank K. Zarembo for useful comments.
We acknowledge financial support from projects  FPA2013-46570  and
 2014-SGR-104.


\appendix

\section{Instantons in ${\cal N}=2^*$}

The  Nekrasov equivariant instanton partition function has the general form
\be
Z_{\rm inst} = \sum_{k=0}^\infty q^k z_k(M,a, \epsilon_1,\epsilon_2)\ ,\qquad q=e^{2\pi i\tau},\ \ \tau=\frac{\theta}{2\pi}+  \frac{4\pi i}{g^2}\ \ .
\ee
In this appendix we compute
the first two coefficients using the construction of \cite{Nekrasov:2002qd,Nekrasov:2003rj}
(see also \cite{Pestun:2007rz,Alday:2009aq}).
We obtain 
\bea
&& z_{1}=
\frac{\left(4 M^2-\left(\epsilon _1-\epsilon _2\right){}^2\right) \left(4 M^2-16 a^2+3 \left(\epsilon _1+\epsilon _2\right){}^2\right)}{8
   \epsilon _1 \epsilon _2 \left(\left(\epsilon _1+\epsilon _2\right){}^2-4 a^2\right)}\ ,
\\
&& z_{2}=\frac{\left(4 M^2-\left(\epsilon _1-\epsilon _2\right){}^2\right)\left(c_0+ c_1 a^2+ c_2 a^4+c_3 a^6 \right) }{256 \epsilon _1^2 \epsilon _2^2 \left(\left(\epsilon _1+\epsilon _2\right){}^2-4 a^2\right) \left(\left(2 \epsilon _1+\epsilon _2\right){}^2-4
   a^2\right) \left(\left(\epsilon _1+2 \epsilon _2\right){}^2-4 a^2\right)}\ ,
\eea
where
\bea
c_0 &=& 64 M^6 \left(8 \epsilon _1^2+17 \epsilon _2 \epsilon _1+8 \epsilon _2^2\right)+16 M^4 \left(40 \epsilon _1^4+301 \epsilon _2 \epsilon _1^3+542 \epsilon _2^2 \epsilon _1^2+301 \epsilon _2^3 \epsilon _1+40 \epsilon _2^4\right)
\nonumber\\
&+&
4 M^2 \left(24 \epsilon
   _1^6+435 \epsilon _2 \epsilon _1^5+1868 \epsilon _2^2 \epsilon _1^4+2978 \epsilon _2^3 \epsilon _1^3+1868 \epsilon _2^4 \epsilon _1^2+435 \epsilon _2^5 \epsilon _1+24 \epsilon _2^6\right)
\nonumber\\
&-&\left(\epsilon _1+\epsilon _2\right){}^2 \left(72 \epsilon
   _1^6-415 \epsilon _2 \epsilon _1^5-3224 \epsilon _2^2 \epsilon _1^4-5730 \epsilon _2^3 \epsilon _1^3-3224 \epsilon _2^4 \epsilon _1^2-415 \epsilon _2^5 \epsilon _1+72 \epsilon _2^6\right)\ ,
\nonumber\\
\nonumber\\
c_1 &=& -512 M^6-128 M^4 \left(37 \epsilon _1^2+86 \epsilon _2 \epsilon _1+37 \epsilon _2^2\right)
\nonumber\\
&-&32 M^2 \left(\epsilon _1+\epsilon _2\right){}^2 \left(67 \epsilon _1^2+446 \epsilon _2 \epsilon _1+67 \epsilon _2^2\right)\nonumber\\
&+&
8 \left(105 \epsilon _1^6-486 \epsilon _2
   \epsilon _1^5-3209 \epsilon _2^2 \epsilon _1^4-5396 \epsilon _2^3 \epsilon _1^3-3209 \epsilon _2^4 \epsilon _1^2-486 \epsilon _2^5 \epsilon _1+105 \epsilon _2^6\right)\ ,
\nonumber\\
\nonumber\\
c_2 &=& 4096 M^4+2048 M^2 \left(5 \epsilon _1^2+13 \epsilon _2 \epsilon _1+5 \epsilon _2^2\right)
\nonumber\\
&-&256 \left(\epsilon _1^2-8 \epsilon _2 \epsilon _1+\epsilon _2^2\right) \left(11 \epsilon _1^2+18 \epsilon _2 \epsilon _1+11 \epsilon _2^2\right)\ ,\nonumber\\
\nonumber\\
c_3&=& -2048 \left(4 M^2-\epsilon _1^2-\epsilon _2^2+8 \epsilon _1 \epsilon _2\right)\ .
\eea
This agrees with the results of \cite{Pestun:2007rz,Billo:2013fi}. Higher instanton coefficients can be
automatically generated from the general formulas of \cite{Nekrasov:2002qd,Nekrasov:2003rj},
but they involve longer expressions.

The computation of the partition function on the four-sphere requires using the Euclidean prescription $a\to i a$ and $M\to iM$ and, in addition,  setting $\epsilon_1=\epsilon_2\equiv \epsilon =1/R$. This gives
\bea
z_{1}&=&- \frac{M^2 \left(4 a^2-M^2+3 \epsilon ^2\right)}{2 \epsilon ^2 \left(\epsilon
   ^2+a^2\right)}
\nonumber
\\
 z_{2} &=& \frac{M^2 }{4 \epsilon ^4 \left(a^2+\epsilon
   ^2\right) \left(4 a^2+9 \epsilon ^2\right)^2} \bigg(64 a^6 \left(2 M^2-3 \epsilon ^2\right)-32 a^4 \left(2 M^4-23 M^2 \epsilon ^2+30 \epsilon
   ^4\right)
\nonumber\\
&+&4 a^2 \left(2 M^6-80 M^4 \epsilon ^2+290 M^2 \epsilon ^4-393 \epsilon ^6\right)
\nonumber\\
&+& 33 M^6 \epsilon
   ^2-306 M^4 \epsilon ^4+477 M^2 \epsilon ^6-804 \epsilon ^8\bigg)
\label{z1z2}
\eea
There is an overall factor of $M^2$ in all instanton contributions. As a result, when $M=0$, all instanton contributions vanish,
as expected, since in this case the theory reduces to ${\cal N}=4$ SYM on ${\mathbb S}^4$.

\section{${\cal N}=2^*$ toy model   without instantons}

In this appendix we study the partition function of ${\cal N}=2^*$ $SU(2)$ theory as a function of the coupling, ignoring instanton contributions.
The model gives some insight on the dynamics of the large $N$ $SU(N)$ gauge theories
in  a simplified context, though, as shown below, instanton contributions are actually important 
at all couplings.

We consider the partition function
\be
Z^{{\cal N}=2^*}= \int da \, e^{-S}\ ,
\ee
where $S$ is given by (\ref{azion}) without the instanton terms.
The saddle-point equation is ($R=1$)
\be
0= \frac{8\pi^2}{g^2} a +  4 a\ln 2|a| -(M+2a)\ln |M+2a|+(M-2a)\ln |M-2a|\ .
\ee
We now solve this equation  in different regimes.

\subsubsection*{Weak $g\ll 1$ coupling regime}

When $g\ll 1$, the saddle point is located at $a\ll M$.
Expanding the saddle-point equation in powers of $a$, we find the solution
\be
\label{sass}
a^*\approx \frac{e}{2} e^{-\frac{2 \pi ^2}{g^2}} M\left(1-\frac{e^2}{6} e^{-\frac{4 \pi ^2}{g^2}}+\frac{7e^4}{360} e^{-\frac{8 \pi ^2}{g^2}}+...\right)\ .
\ee
Substituting into the action, we find the free energy $F=-\ln Z$,
\be
F=  R^2M^2
 \Big( 2\ln M + e^{2} e^{-\frac{4 \pi ^2}{g^2}}  -\frac{e^4}{6}\ e^{-\frac{8 \pi ^2}{g^2}}
-\frac{e^6}{30} \ e^{-\frac{12 \pi ^2}{g^2}}-
\frac{e^8}{84} e^{-\frac{16 \pi ^2}{g^2}} 
+...\Big)\ .
\ee
This is a non-perturbative expansion which is not  due to instantons, which are not incorporated in this toy model. The physical origin of such terms is, as in the SQCD model, the OPE expansion. Upon including instantons, there will be a mixing of non-perturbative terms
of different origin. For an $SU(N)$ gauge group with  large $N$, there is no mixing, because instantons are suppressed like  $ e^{-\frac{8\pi^2N}{\lambda}}$, whereas OPE contributions are finite contributions of
order $ e^{-\frac{8\pi^2}{\lambda}}$.

Note that  truncating the instanton expansion (\ref{variat})  is justified if $g\ll 1$ and  the saddle point lies on a region
\be
\label{condis}
a \gg M e^{-\frac{2\pi^2}{g^2}}\ .
\ee
The location of the saddle point (\ref{sass}) does not satisfies this condition.
Therefore instantons cannot be ignored in this regime.

\subsubsection*{ Phase transition at $g\approx g_{\rm cr}$}

As $g$ is gradually increased from 0, the value of $a_*$ monotonically increases until a critical value where $a_{\rm cr}^*= M/2$.
This occurs when
\be
\label{tras}
 e^{-\frac{2 \pi ^2}{g^2_{\rm cr}}} = \frac{1}{2}\ .
\ee
 Just below $g_{\rm cr}$, $a_*$ exhibits a non-analytic behavior defined by
 \be
\frac{4\pi^2}{g_{\rm cr}^4}  (g^2-g^2_{\rm cr})\approx \frac{1}{M} (M-2a_*)\ln(M-2a_*) \ .
 \ee
Again we note that neglecting instantons is not justified,
since the condition (\ref{condis}) is not satisfied near $a^*=M/2$, see (\ref{tras}).

 \subsubsection*{Strong  $g\gg 1$ coupling regime}

Expanding $S$ at large $a$, we find that the saddle point is at
 \be
 a_* =M \frac{g}{4\pi}\ .\ \
 \ee
Interestingly, this is of a similar form as the formula found for the width of the eigenvalue distribution for $SU(N)$ (in that case, the width was given by $M\sqrt{g^2N}/2\pi$).

 Substituting into the action, we find the free energy in the strong coupling limit $g\gg 1$,
 \be
 F\cong - M^2 R^2 \ln \left(g^2 M^2 R^2\right)\ .
\ee

\end{document}